%% file: main.tex
\documentclass[sigconf]{acmart}
\AtBeginDocument{%
  \providecommand\BibTeX{{%
    \normalfont B\kern-0.5em{\scshape i\kern-0.25em b}\kern-0.8em\TeX}}}

\copyrightyear{2022} 
\acmYear{2022} 
\setcopyright{acmlicensed}\acmConference[CHI '22]{CHI Conference on Human Factors in Computing Systems}{April 29-May 5, 2022}{New Orleans, LA, USA}
\acmBooktitle{CHI Conference on Human Factors in Computing Systems (CHI '22), April 29-May 5, 2022, New Orleans, LA, USA}
\acmPrice{15.00}
\acmDOI{10.1145/3491102.3502141}
\acmISBN{978-1-4503-9157-3/22/04}

\usepackage{subcaption}
\usepackage[capitalize, noabbrev]{cleveref}

\usepackage{calc}

\definecolor{HaisColor}{rgb}{0.9,0.3,0.9}

\definecolor{LukasColor}{rgb}{0,0.3,0.9}

\definecolor{DanielsColor}{rgb}{0.9,0.6,0.1}

\definecolor{MKColor}{rgb}{0.1,0.6,0.1}

\newcommand{\glmmci}[5]{$\beta$=#1, SE=#2, CI$_{95\%}$=[#3, #4], p#5}

\newcommand{\ttest}[7]{mean diff. = #1, CI$_{95\%}$=[#2, #3], t(#4) = #5, p#6; d = #7}

\newcommand{\lastaccessed}{\textit{last accessed 01.12.2021}}

\newcommand{\pct}[1]{#1\,\%}

\newcommand{\ivslidertype}{\textsc{Type}}
\newcommand{\ivnumber}{\textsc{Number}}

\newcommand{\slidertyperegular}{\textit{Regular}}
\newcommand{\slidertypefilmstrip}{\textit{Filmstrip}}

\newcommand{\prestudyN}{12}
\newcommand{\mainStudyNraw}{156}
\newcommand{\mainStudyN}{138}

\begin{document}

\title[GANSlider: How Users Control Generative Models for Images]{GANSlider: How Users Control Generative Models for Images using Multiple Sliders with and without Feedforward Information}

\author{Hai Dang}
\email{hai.dang@uni-bayreuth.de}
\affiliation{%
  \institution{University of Bayreuth}
  \city{Bayreuth}
  \country{Germany}
}

\author{Lukas Mecke}
\email{lukas.mecke@unibw.de}
\affiliation{%
  \institution{Bundeswehr University Munich}
  \city{Munich}
  \country{Germany}
}
\affiliation{%
  \institution{LMU Munich}
  \city{Munich}
  \country{Germany}
}

\author{Daniel Buschek}
\email{daniel.buschek@uni-bayreuth.de}
\affiliation{%
  \institution{University of Bayreuth}
  \city{Bayreuth}
  \country{Germany}
}

\renewcommand{\shortauthors}{Dang, et al.}

\begin{abstract}

We investigate how multiple sliders with and without feedforward visualizations influence users' control of generative models. In an online study (N=138), we collected a dataset of people interacting with a generative adversarial network (\textit{StyleGAN2}) in an image reconstruction task. We found that more control dimensions (sliders) significantly increase task difficulty and user actions. Visual feedforward partly mitigates this by enabling more goal-directed interaction. However, we found no evidence of faster or more accurate task performance. This indicates a tradeoff between feedforward detail and implied cognitive costs, such as attention. Moreover, we found that visualizations alone are not always sufficient for users to understand individual control dimensions. Our study quantifies fundamental UI design factors and resulting interaction behavior in this context, revealing opportunities for improvement in the UI design for interactive applications of generative models. We close by discussing design directions and further aspects.

\end{abstract}



\begin{teaserfigure}
  \includegraphics[width=\textwidth]{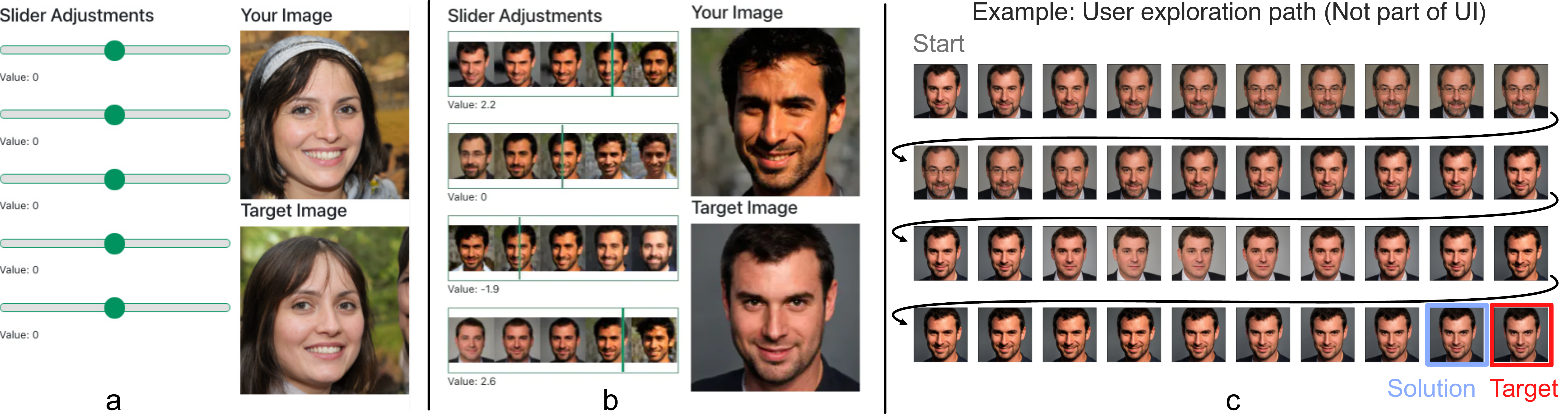}
  \caption{This paper compares different user interfaces with sliders for interactive control of a generative model for face images (\textit{StyleGAN2}). We study different numbers of sliders and two designs: (a) \slidertyperegular{} sliders and (b) \slidertypefilmstrip{} sliders that provide feedforward information via preview images. Participants used these UIs in image reconstruction tasks to generate a given target images. The right part of the figure (c) shows one example of a resulting ``exploration path'' towards a target (for more examples see \cref{sec:appendix}).}
  \Description{Overview of the user study UI with different slider variants and multiple sliders.}
  \label{fig:teaser}
\end{teaserfigure}

\maketitle

\input{sections/introduction}
\input{sections/related_work}

\input{sections/concepts}

\input{sections/user_study}

\input{sections/results}
\input{sections/discussion}
\input{sections/conclusion}

\begin{acks}
We thank Christina Schneegass, Markus Klar, Pascal Knierim and Martin Zürn for their feedback on the manuscript. This project is funded by the Bavarian State Ministry of Science and the Arts and coordinated by the Bavarian Research Institute for Digital Transformation (bidt).
\end{acks}

\bibliographystyle{ACM-Reference-Format}
\bibliography{references}

\input{sections/appendix}

\end{document}

%% file: sections/introduction.tex
\section{Introduction}

Artificial intelligence (AI) supported interactive tools are now being adopted by both expert and non-expert users alike. Especially generative machine learning (ML) models find their application in end user products where they enable users to generate realistic media. In this paper we focus on the use case of image generation, such as seen in the recently introduced \textit{Smart Neural Filters} in Photoshop~\cite{tack_adobe_2020}. These allow users to change semantic features of face portrait photos. 

The underlying models are typically trained on large unlabeled image datasets to learn data representations useful for generation and manipulation of photos. %
This learned representation is called latent space and is often high-dimensional and not always interpretable by humans. However, there are techniques to facilitate the learning or extraction of \textit{disentangled} dimensions~\cite{Bengio2013, ganspace, Ridgeway2016}, such that each one represents a (more) interpretable feature. For example, for faces these might be hair color, eye gaze direction, and so on.

End user products can invest in engineering efforts to achieve clearly interpretable dimensions (cf. the Photoshop example~\cite{tack_adobe_2020}, GauGAN~\cite{gaugan2019}). In contrast, such disentangled dimensions are not always readily available to (AI) researchers while developing new generative models. As a new way to handle this, researchers have worked on methods to explore and evaluate generative models via interaction~\cite{ganspace, Ross2021}. %
This often involves visualizing a model's learned dimensions to judge their interpretability and quality: A popular method in (AI) research publications is to plot multiple images generated at different points along said dimensions (e.g. \cite{ganspace, Karras_2019_CVPR, Karras_2020_CVPR, park2020swapping}). Recently, such image grids have also been used interactively (e.g.~\cite{park2020swapping, Zhang2021}).

Today, most user interfaces (UIs) for generative models use a set of regular sliders, in which each one is mapped to the control of one latent dimension. Such UIs have been used to enable interaction for model exploration and evaluation \cite{ganspace, Ross2021, park2020swapping}. However, slider UIs \textit{themselves} have not been at the focus of any empirical study in this context so far.  %

This motivates us to better undestand the interaction patterns with sliders for the control of generative models for images. Concretely, we hypothesize that here the design of effective UIs is constrained by different factors: 
First, we investigate how the \textit{number of control dimensions} (i.e. sliders) impacts the interaction. An increasing number of control dimensions corresponds to a larger latent image space that can be explored, and thus adds to the complexity of interactive tasks in that space. Currently, there is no guidance on the number of control dimensions to use in various settings, with numbers reported in related work ranging from 5 to 80 dimensions \cite{ganspace, Ross2021, park2020swapping}. 
Second, we hypothesize that \textit{visual feedforward information} in the UI may aid users in their decision-making when working with %
generative models (i.e. a preview of the outcome of performing an interaction). This motivates us to introduce and evaluate a ``filmstrip'' slider design, which shows multiple preview images generated at different points along the controlled dimension (\cref{fig:teaser} b). %
This leads to the following two research questions that we examine in this paper:

\begin{itemize}
    \item [\textbf{RQ1}] How do different \textit{numbers of control dimensions} (sliders) impact users' interaction behaviour and reconstruction task performance with a generative image model?
    \item [\textbf{RQ2}] How does added \textit{visual feedforward information} (preview images) on the sliders impact users' interaction behaviour and reconstruction task performance with a generative image model?
\end{itemize}

To answer these questions, we conducted a within-subject user study (N=\mainStudyN) in which each participant worked with $\{$1, 2, 3, 4, 5, 8, 10$\}$ sliders of both types (\slidertyperegular{}, \slidertypefilmstrip{}). We used an image reconstruction task (cf. \cref{fig:teaser}) similar to \citet{Ross2021}. We measured all interaction events plus subjective feedback. %

Our results show that the number and type of sliders both significantly influence the interaction: More sliders lead to more interactions and add to the complexity of the task, which filmstrips partly mitigate. However, we found no evidence of faster or more accurate task performance with filmstrips. Furthermore, we observed that controlling generative models can already become challenging for up to ten sliders. While our results indicate that filmstrips alone are not always sufficient to interpret a model's latent dimensions, they do lead to more goal-oriented user actions and were overall well-received by the participants. Furthermore, we found that it is crucial to look beyond performance measures such as speed and accuracy to evaluate UI designs in this context. Based on these results and the collected open feedback we discuss implications for UIs for generative models. %

In summary, we contribute: (1) The first empirical evaluation of two fundamental UI design factors for interaction with generative models; involving (2) a new slider design with ``filmstrip'' previews as feedforward information; and (3) the resulting dataset and prototype, which we release to the community to facilitate further research.

%% file: sections/related_work.tex
\section{Background and Related Work}

\begin{figure*}
    \centering
    \includegraphics[width=0.9\linewidth]{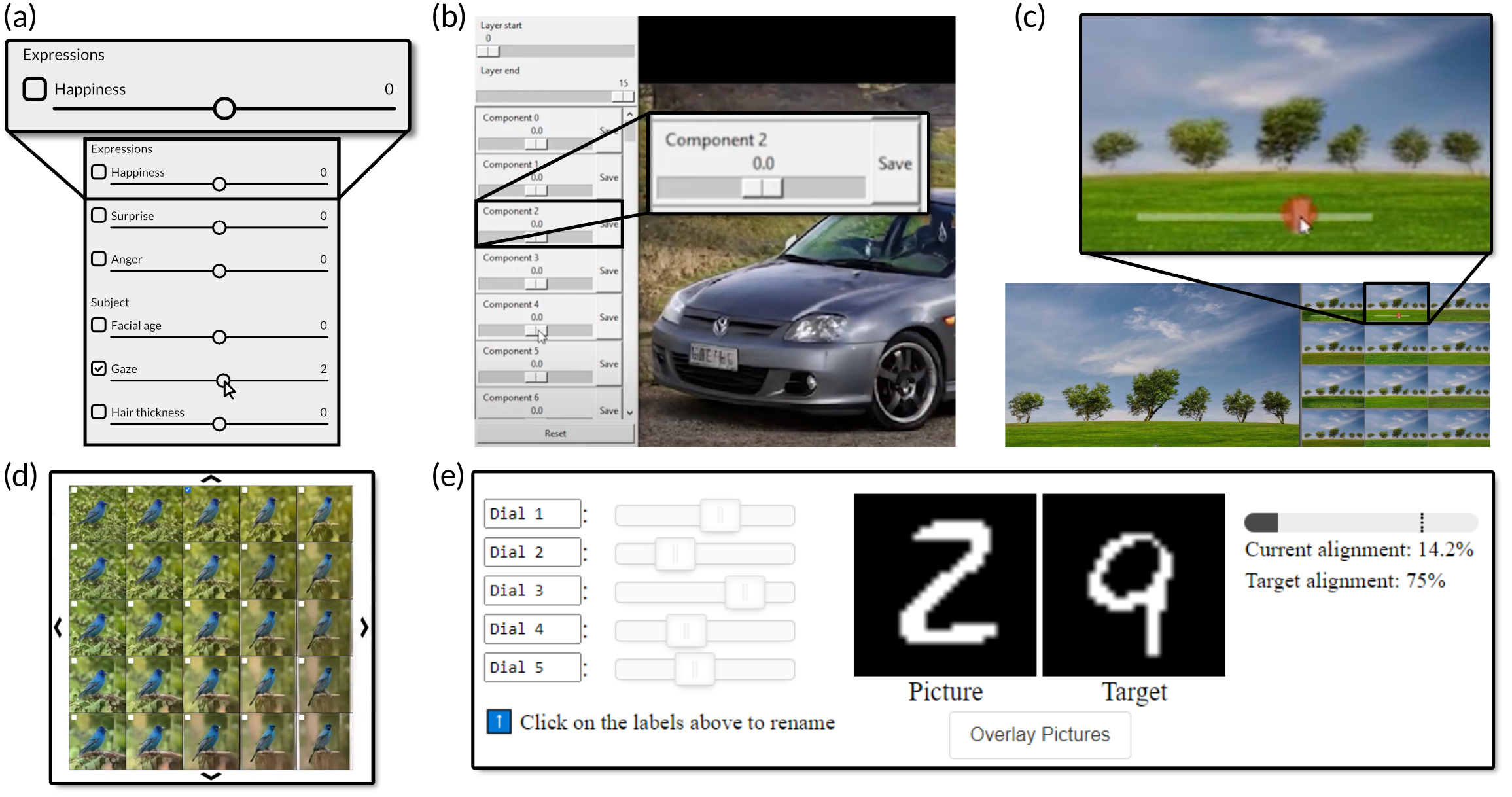}
    \caption{Examples of the many slider and image grid UIs used in related work to interact with generative image models: \textit{(a)} UI for editing face photos with the ``Smart Portrait Filters'' in Adobe's Photoshop~\protect\cite{tack_adobe_2020}; \textit{(b)} UI used in the \textit{GANSpace} paper~\protect\cite{ganspace} (sliders manipulate latent dimensions discovered via PCA); \textit{(c)} UI used in the ``Swapping Autoencoders'' paper~\protect\cite{park2020swapping} (thumbnail grid; sliders on thumbnails change strength of the corresponding texture edits); \textit{(d)} Automatically created image gallery for interactive GAN exploration (image grid without sliders)~\protect\cite{Zhang2021}; \textit{(e)} slider UI used in a user study to evaluate model interpretability by interactive reconstruction~\protect\cite{Ross2021}. In this paper, we examine a new combination of sliders and images for controlling generative models -- ''filmstrip sliders'' (cf.~\protect\cref{fig:teaser}). Image sources: Screenshots from the related work's videos/websites.\protect\footnotemark Callouts added for this figure.}
    \label{fig:slider_uis_in_related_work_compilation}
    \Description{
        Overview of five different UI designs for working with generative models. Four designs use slider elements in the UI and one design uses a grid of images for the exploration of the generative model.
    }
\end{figure*}
\footnotetext{\textit{(a) }Mockup by the authors, following the Photoshop UI shown in \url{https://youtu.be/5rIl2-Gw8lQ}; \textit{(b)} \url{https://youtu.be/jdTICDa_eAI}; \textit{(c)} \url{https://youtu.be/0elW11wRNpg}; \textit{(d)} \url{https://youtu.be/5hgLDjtvlBg}; \textit{(e)} \url{http://hreps.s3.amazonaws.com/quiz/manifest.html}; all \lastaccessed.}

We structure this related work section into three parts: a motivation for interaction with generative models (\cref{sec:related_work_motivation}), frontend (i.e. UI, HCI view -- \cref{sec:related_work_uis_for_gen_models}) and backend (i.e. modeling, AI view -- \cref{sec:related_work_model_dimensions}). %

\subsection{Motivation for Studying Interaction with Generative Image Models}\label{sec:related_work_motivation}

Our work is motivated by the growing relevance of interactive generative models in both research, industry and products (e.g. Artbreeder\footnote{\url{https://www.artbreeder.com/}, \lastaccessed.},
GauGAN~\cite{gaugan2019},
Photoshop~\cite{tack_adobe_2020}, \cref{fig:slider_uis_in_related_work_compilation} a). This trend is expected to continue, given the investment of major tech companies~\cite{scott_microsoft_2020, tack_adobe_2020, Wright2020}.
Recent research in Human Computer Interaction (HCI) and AI used generative models interactively as well (e.g. see~\cite{ganspace, PoursabziSangdeh2021, Ross2021} and \cref{fig:slider_uis_in_related_work_compilation}), including work on co-creativity and computational creativity (e.g.~\cite{Rafner2020chiplay, Rafner2021cc}). %
The study by \citet{Ross2021} is most closely related: 
They evaluated the interpretability of such models by letting people reconstruct a target image with sliders to control the model (\cref{fig:slider_uis_in_related_work_compilation} e). 
We follow their approach and study image reconstruction with sliders. However, while their goal was to evaluate the models, our goal is to evaluate the slider UI. %
We see these goals as supporting each other since understanding user behaviour is crucial for further establishing interactive tasks as an evaluation method for AI: For instance, UIs for model comparisons need to be well-understood to avoid bias. Evaluation tasks should also elicit realistic behaviour, and in the context of interactive AI ``researchers need to be cautious about their pragmatic decisions''~\cite{Buccinca2020}. %
So far, the regular slider UI is such a pragmatic choice for generative models for two reasons: First, slider design and use has not been studied in this context yet. Second, comparisons to alternative UI elements are missing. In this paper, we focus on the first aspect -- better understanding the slider design and interaction behaviour.

\subsection{Frontend/UI: User Interfaces for Generative Models}\label{sec:related_work_uis_for_gen_models}

\subsubsection{Slider UIs}

The \textit{slider} is the most commonly found UI element for interaction with generative models today \cref{fig:slider_uis_in_related_work_compilation} shows examples from research and industry. Typically, each slider maps to a shift along one latent dimension (e.g. \cref{fig:slider_uis_in_related_work_compilation} a, b, c, e). These dimensions could be predetermined (e.g. in end user applications such as in \cref{fig:slider_uis_in_related_work_compilation} a: ``happiness'', ``gaze'', etc.) or open for discovery (e.g. in research contexts such as \cref{fig:slider_uis_in_related_work_compilation} b and e).
Sometimes, sliders are accompanied by images that visualize the dimension (e.g. \cref{fig:slider_uis_in_related_work_compilation} c) and/or the overall outcome (e.g. \cref{fig:slider_uis_in_related_work_compilation} a, b and e).
However, interaction with sliders with and without images has not been systematically evaluated yet for generative models. Without this baseline, it is difficult to advance the UI design for this use case. %
This motivates us to investigate how well the basic slider design performs and how it might be improved with visuals.

\subsubsection{Other UIs}

2D grids are another presentation for image generation models: For example, \citet{Zhang2021} generate galleries to explore Generative Adversarial Networks (GANs) (\cref{fig:slider_uis_in_related_work_compilation} d). This affords navigation such as selecting the next image on which to center the grid. Typically, grids map x/y axes to two directions in a model's latent space, showing images at equidistant points along those. Non-interactive variants frequently appear in (AI) publications to visualize (1) latent spaces (e.g.~\cite{ganspace}), (2) the impact of varying inputs and hyperparameters (e.g.~\cite{park2020swapping, Karras_2019_CVPR}), or (3) generated examples (e.g.~\cite{Karras_2020_CVPR}). %
Our filmstrip design explores a combination of sliders with such image concepts.

Further UI concepts edit image parts or aspects: \citet{park2020swapping} enable structural/textural edits via thumbnails and sliders to control their strength (\cref{fig:slider_uis_in_related_work_compilation} c).
Others allow users to mark image areas to (not) be modified by the model~\cite{yang2020mask_guided}, to guide the model~\cite{Turmukhambetov2015, zhu2016generative}, or to specify semantic edits (e.g. add windows, remove chairs)~\cite{Bau2019semanticImagePrior} .
In this paper, we focus on sliders for global image edits to analyze interactions in-depth. %
However, localized edits (e.g. brushing, pen tools) could be studied with similar methods in the future (e.g. logging interactions in local editing tasks).

\subsubsection{Extending Sliders with Feedforward}

Previews~\cite{Schwarz2015} and feedforwad concepts~\cite{Bau2008, Coppers2019} inform users about the outcome of their actions before they complete them. The HCI literature includes related slider designs: \citet{Willett2007} added visual ``scents'' to navigate datasets, for example, with a slider with an added bar chart. Related, \citet{SketchSlider} enabled users to draw sliders, with density plots added automatically along the drawn line. Similarly, \citet{Kwon2017} added ``rainbow'' visualizations of data attributes along user-defined, non-linear axes of a scatterplot. \citet{TerrySideView} proposed ``side views'' -- pop ups that show visual previews of the states of a UI element to help users explore its functionality. %
Our filmstrip slider builds on these ideas of adding visual (preview) information to the slider line. %

\subsection{Backend/Modeling: Generative Models and Selecting Dimensions}\label{sec:related_work_model_dimensions}

ML/AI research has improved the quality of generated output and its scope of applications (e.g. text and images) with increasing model sizes~\cite{brown2020language} and refined training and architectures~\cite{Karras_2019_CVPR, Karras_2020_CVPR, Karras2021, qiu_pre-trained_2020}. Motivated by creating models that can be understood by humans, efforts were taken to make them more interpretable and to measure interpretability~\cite{Ross2021}. Here, the term \textit{disentanglement} is used to describe how ``clear cut'' the dimensions of the learned data representations are and how well they match the underlying ``true'' factors~\cite{Bengio2013, Ridgeway2016}. This is also important for interactive use: For example, for face images, an entangled dimension might change both hairstyle and skin tone, while a disentangled model might have learned these as two separate dimensions. 

As described in \cref{sec:related_work_motivation}, \citet{Ross2021} compared models with varying degrees of disentanglement using an interactive task with one UI. We instead compare varying UI designs and thus chose one model, \textit{StyleGAN2}~\cite{Karras_2020_CVPR}, trained on face images. This is motivated by exploring a task that goes beyond more abstract or simpler datasets (e.g. Sinelines, MNIST)~\cite{Ross2021}. Portrait editing with generative models is also practically relevant for both end users and researchers: It has been added to recent products~\cite{tack_adobe_2020} and is (inter)actively investigated in ongoing AI research~\cite{ganspace, Karras_2019_CVPR, Karras_2020_CVPR, Karras2021}. 

To obtain interesting dimensions to be controlled in the UI, we can either facilitate learning disentangled representations (e.g.~\cite{Bengio2013}) or ``discover'' meaningful concepts in a model: %
For our study, we used the \textit{GANSpace} approach~\cite{ganspace} to automatically select the dimensions from \textit{StyleGAN} to be controlled via sliders in our study (\cref{sec:apparatus_web_system}), without manual labeling. 
Our motivation here is a middle ground between random (and thus likely meaningless) dimensions and fully ``engineered'' interpretable dimensions. The latter are important for end user applications (e.g. the Photoshop example in \cref{fig:slider_uis_in_related_work_compilation} a) but cannot be readily expected in earlier stages (e.g. research and development). Moreover, related work has recently identified interaction as a key factor for evaluating generative models that are not perfectly disentangled~\cite{PoursabziSangdeh2021, Ross2021}. At the same time, the UI has not been in focus here yet, further motivating our study.

%% file: sections/concepts.tex
\section{Extending the Slider UI with Filmstrips}

As one of our variables of interest, we examine the impact of adding further visual information to the sliders (see \cref{fig:teaser}). Here, we describe this UI concept in more detail.

\subsection{Filmstrip Previews}

In AI research, image strips/grids are widely used to present generative model results (e.g. \cite{ganspace, Karras_2019_CVPR, Karras_2020_CVPR, park2020swapping}). %
This is mostly non-interactive (e.g. paper figures). In HCI and information visualization, previews and feedforward concepts~\cite{Bau2008, Coppers2019, Schwarz2015, TerrySideView, Willett2007} support interaction and ``what if'' reasoning in tasks. Our filmstrip concept combines these experiences (\cref{fig:teaser}). %
We chose a fixed number of images for visual orientation along the underlying continuous control dimension. Concretely, we show five images per slider, as informed by a pretest with N=12 people, the available space on a regular desktop screen, and considering computational costs. 
Slider ranges were set to [-5, 5]. In our prestudy, we found this to be a good compromise between expressiveness and robustness, since, as typical for models such as \textit{StyleGAN}, moving further along a dimension causes stronger changes but can eventually lead to visual artefacts (cf.~\cite{ganspace}). 

\subsection{Linked Sliders}

The output image of the generative model is the sum of all slider edits. If one slider changes the hair color, the color in all filmstrips has to be updated if we want the filmstrip on each slider to always show the concrete outcome of changing that slider. An alternative design might show a generalised visual representation per slider, such as a filmstrip with different hair colors (e.g. color swatches or a fixed face with different hair colors). However, this approach assumes that the dimensions have known and fixed interpretations. Thus we decided for the first option here. Accordingly, in our UI, moving one slider updates the filmstrip preview images of all sliders.

\subsection{Side-by-Side Outcome View}

We include an \textit{outcome view} with two images (``Your Image'' and ``Target Image'' in \cref{fig:teaser}), following related work, where people found this side-by-side view more helpful than overlays~\cite{Ross2021}. %
Beyond research studies, this design is also relevant for applications: For example, for image editing, one image could show the original and one the edited version.

\subsection{Technical Implementation Aspects}\label{sec:technical_implementation}

The preview images are computationally costly because they require to run \textit{StyleGAN} for all sliders after each interaction. To address this, we  %
implemented the following update scheme: 
With each slider change event the client requests an update from the server. This also aborts all pending requests to avoid spending computation time on now outdated events. The server computes and sends back the image(s) or an empty response if no computing time is available. In case of an empty response, the client retries immediately and keeps the previous image until the update succeeds. This effectively realises that updates come as fast as possible for the available computing power and changing sliders is always possible (non-blocking UI). Concretely, a slider was greyed out as a whole while any of its images was updating. This was informed by preliminary testing, to clearly indicate outdated previews and ongoing computation, while also reducing the amount of UI ``flickering'' that this necessarily introduces. Moreover, the outcome image (``Your Image'', \cref{fig:teaser}) had update priority.

%% file: sections/user_study.tex
\section{User Study}

We conducted a user study to investigate how different numbers and types of sliders impact users' interaction behaviour and reconstruction task performance with a generative image model.

\subsection{Study Design}

We designed a within-subject online study with two independent variables: the sliders' \ivslidertype{} (two levels: \slidertyperegular{} and \slidertypefilmstrip) and the \ivnumber{} of sliders (seven levels: 1, 2, 3, 4, 5, 8, 10; informed by pretests that showed 10 to be clearly challenging). As dependent variables, we computed interaction measures from logging data, and elicited feedback via questionnaires (Likert items and open questions).

\subsection{Apparatus}\label{sec:apparatus}

\subsubsection{Web System}\label{sec:apparatus_web_system}
We implemented a light server for message passing (flask\footnote{\url{https://flask.palletsprojects.com/en/2.0.x/}, last visited \today}) and a client (ReactJS\footnote{\url{https://reactjs.org/}, last visited \today}). Besides the sliders (\cref{fig:teaser}), this frontend included a study consent form, task descriptions, done/skip buttons, and questionnaires.

The model ran on a multi-GPU instance on AWS Sagemaker\footnote{\url{https://aws.amazon.com/sagemaker/}, last visited \today}. %
We used the \textit{StyleGAN2} model~\cite{Karras_2020_CVPR} %
trained on the FFHQ dataset~\cite{Karras_2019_CVPR}
and applied the \textit{GANSpace} approach (i.e. Principal Component Analysis, PCA, on \textit{StyleGAN}'s $\mathbf{w}$ vectors, see~\cite{ganspace} for details): %
Concretely, for a task with N sliders, we use this to extract the N top dimensions as ranked by PCA. %
We then map this ranking to the UI layout (i.e. first slider controls first principal component). 
We chose this ranked extraction approach as a balance between random control dimensions and manually selected or ``engineered'' ones (see \cref{sec:related_work_model_dimensions}). %

In contrast to \citet{Ross2021}, we do not show numerical real-time feedback (e.g. remaining distance to target image) since they reported users sometimes used this distance as the only source of information to solve the task and thereby did not pay attention to the sliders.

\subsubsection{Generation of the Image Reconstruction Tasks}
\textit{StyleGAN2}~\cite{Karras_2020_CVPR} is capable of generating ``random'' faces, which we use in this study. %
For each task, the starting face and target face are defined as follows: First, we choose a random point (i.e. a face) in \textit{StyleGAN2's} latent space. This is the \textit{starting image} for the task shown with all sliders at the center (i.e. at 0). Second, we choose a random target offset for each slider. Since each slider corresponds to a direction in the latent space, these values define an offset in the latent space.
Third, we shift the point by this offset and use \textit{StyleGAN2} to generate the \textit{target image}.  
As an example, for a task with only one slider, we might roll an offset of ``-2.2''. In this case, the task can be solved perfectly by moving the slider from its initial position (center) at zero to ``-2.2'' (towards the left).

We created a set of seven random faces for each participant. These faces were randomly assigned to each task for each slider variant, such that the seven tasks (i.e. numbers) for both slider variants used all seven faces, but in a different order. This approach was chosen to ensure these three aspects: (1) The study overall covers a wide range of model outputs (random faces across participants); (2) a potential influence of specific faces on task difficulty is balanced because for one participant the same faces appear in both slider variants; (3) it is still not possible for participants to know the correct slider configuration for a face when encountering it again in the second variant, as it %
appears with different target offsets (i.e. target slider values) and a different number of sliders.

\subsubsection{Task Questionnaires}\label{sec:study_questionnaires}
We used the NASA-TLX questionnaire \cite{NASATLX} to measure workload and four custom Likert statements (see titles in \cref{fig:likert}). %
These were presented directly in the web system after each reconstruction task.

\subsubsection{Final Questionnaire}\label{sec:study_final_questionnaire}
A final questionnaire asked for open feedback: \textit{Do you have any final remarks on the interpretability of the different slider types?}, \textit{Is there anything you would add to the slider interface to improve it?}, and \textit{Can you think of other use cases where the filmstrip slider interface would be helpful?} These questions were not mandatory, to account for the case that participants had no feedback/ideas.

\subsection{Participants}

We recruited \prestudyN{} participants for a pilot study and \mainStudyNraw{} for the main study, using \textit{Prolific}\footnote{\url{www.prolific.co}, \lastaccessed}. 
The study took 45 minutes on average and was compensated with \pounds\,5.63 (\textit{Prolific} uses \pounds), following \textit{Prolific's} recommendation. %
After exclusions (e.g. incomplete or invalid attempts, such as not moving the sliders), we had N=\mainStudyN{} participants for the analysis.

\subsection{Procedure}

\subsubsection{Study Intro}
In line with our institutional regulations and informed consent procedures, the first page explained the study, provided detailed information on data collection and privacy protection regulations, and further general study information (e.g. emphasizing that participants could end the study at any time). Before the first actual task, an introductory page allowed participants to try out single sliders of both variants. %

\subsubsection{Image Reconstruction Tasks}
Each participant completed 14 image reconstruction tasks: 
They first completed one set of seven tasks corresponding to one slider \ivslidertype{} before switching to the other one. This was counterbalanced (i.e. half of participants started with the seven tasks for \slidertyperegular, the others with the seven tasks for \slidertypefilmstrip). 

The order of these seven tasks per slider variant was fixed with the \ivnumber{} of sliders increasing. This decision was informed by our prestudy, which revealed that tasks with higher slider numbers can be very challenging. %
In this way, a potential learning effect in our study is in favour of participants being able to solve the tasks with many sliders. %
Similar study designs with tasks of increasing interaction difficulty have been used successfully in related work~\cite{Mecke2019soups}. %

\subsubsection{Ending a Task}\label{sec:ending_a_task}
We did not use a time limit or a predefined similarity threshold for ending a task. Instead, people had two options to end a task: They could either indicate that they had successfully finished it (``done'') or that they wanted to ``skip'' it. We did this to allow users to fully concentrate on solving the task to the best of their ability and assess their perceived success. %
Each task was followed by the questionnaires described in \cref{sec:study_questionnaires}.

\subsubsection{Ending the Study}
The final questionnaire (\cref{sec:study_final_questionnaire}) concluded the study with open feedback.

%% file: sections/results.tex
\section{Results}

Here we report the results of our study.
Where applicable, we report significance at p~<~.05, tested with R~\cite{R2020}: We use (generalised) linear mixed-effects models (LMMs, packages \textit{lme4}~\cite{Bates2015} and \textit{lmerTest}~\cite{Kuznetsova2017}). These LMMs accounted for individual differences using random intercepts (for participant), and had \ivslidertype{} and \ivnumber{} as fixed effects.
For Likert (ordinal) data, we use Generalized Estimating Equations (GEEs) from the R package \textit{multgee}~\cite{multgee2015}. 
Where not stated otherwise, we exclude the data of tasks that participants marked as skipped (see \cref{sec:ending_a_task}). %

\begin{figure*}
    \centering
    \begin{subfigure}{0.45\linewidth}
        \centering
        \includegraphics[width=\linewidth]{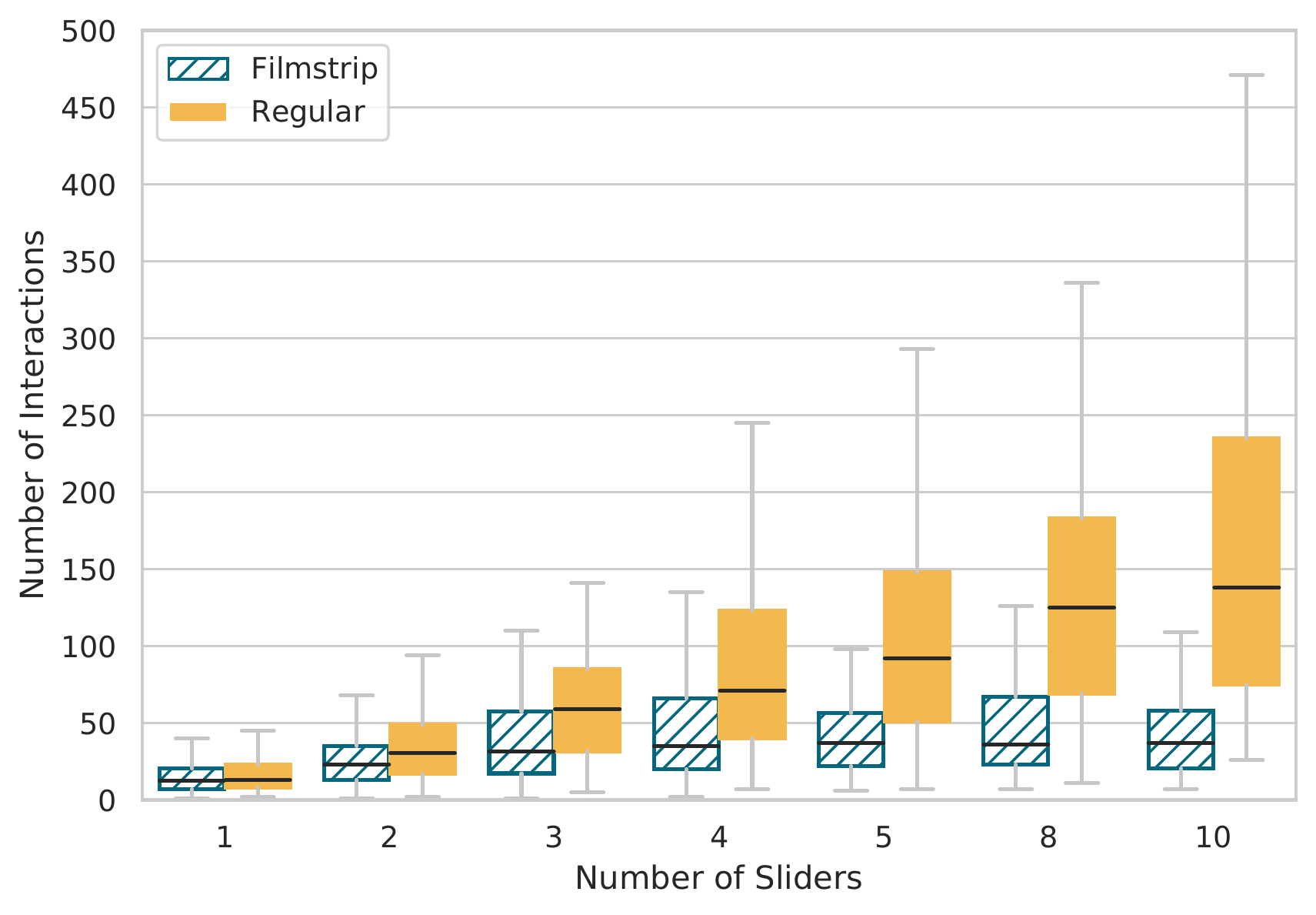}
        \subcaption{Number of interactions}\label{fig:interactions_number}
        \Description{
            Figure shows the total number of interactions for each task. The number of interactions initally increase for the filmstrip variant until four sliders and decreases afterwards. For the regular variant the number of interactions increase with increasing number of sliders. Overall, the median number of interactions for the regular variant is significantly higher than it's corresponding filmstrip variant. 
        }
    \end{subfigure}
    \begin{subfigure}{0.45\linewidth}
        \centering
        \includegraphics[width=\linewidth]{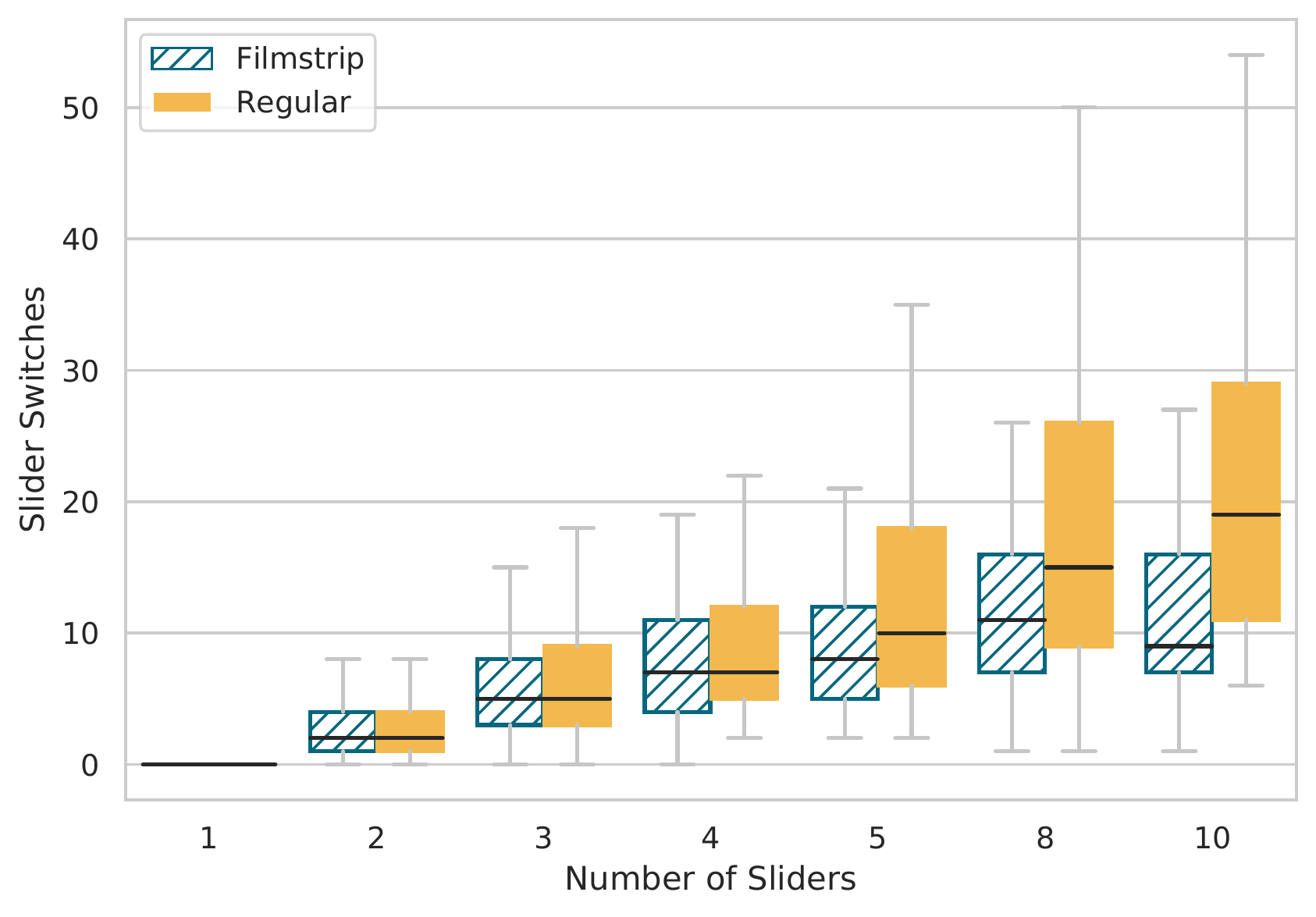}
        \subcaption{Number of control dimension (slider) switches}\label{fig:interactions_slider_switches}
        \Description{
            The number of slider switches increase with any additional slider for both filmstrip and regular. Overall, the number of slider switches for regular is higher than for filmstrip.
        }
    \end{subfigure}
    \begin{subfigure}{0.45\linewidth}
        \centering
        \includegraphics[width=\linewidth]{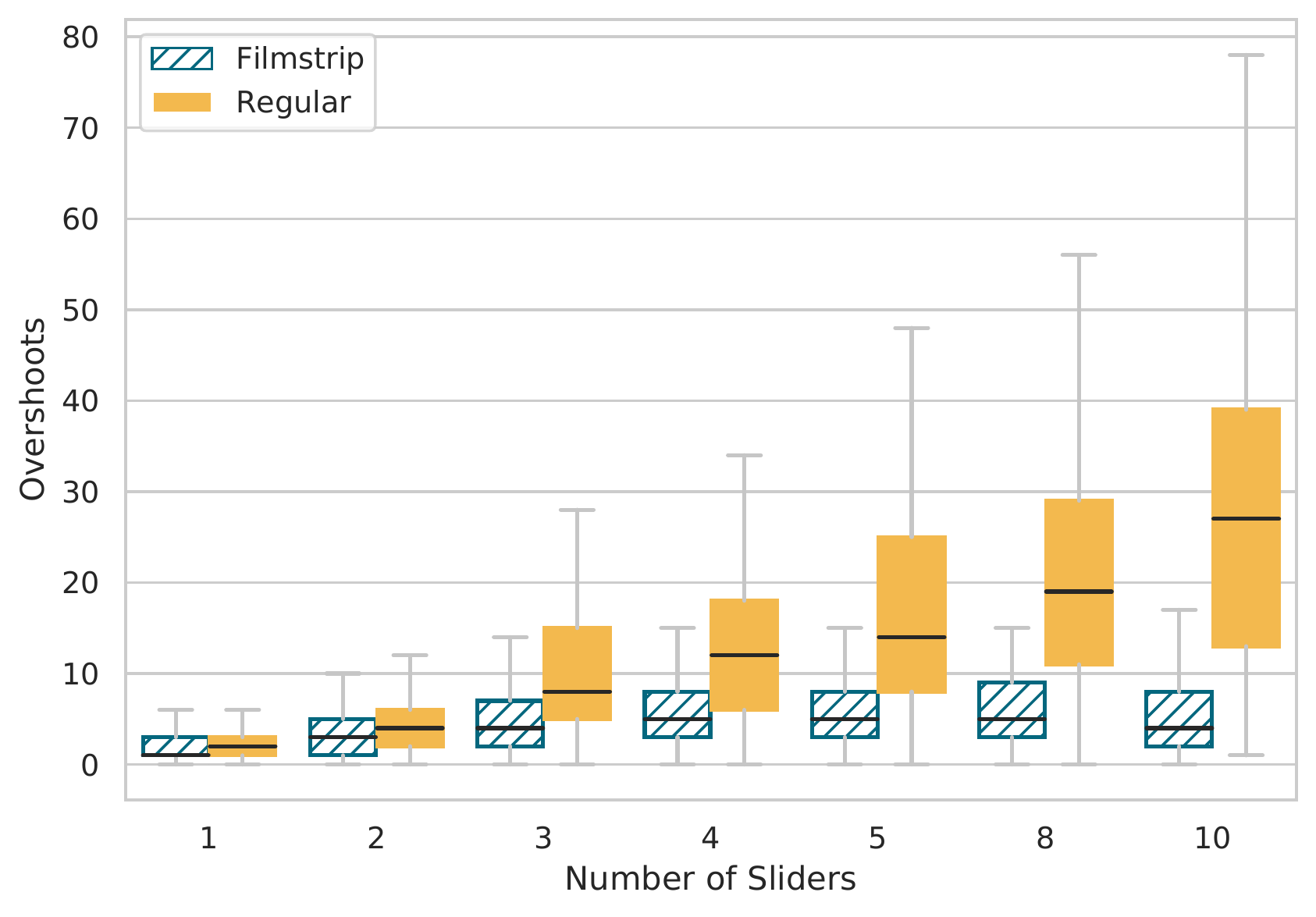}
        \subcaption{Number of overshooting actions.}\label{fig:interactions_overshoots}
        \Description{
            The number of overshoots for the filmstrip variant marginally increase with additional sliders. For the regular variant the number of overshoots over the target value monotonically increases with additional sliders.
        }
    \end{subfigure}
    \begin{subfigure}{0.45\linewidth}
        \centering
        \includegraphics[width=\linewidth]{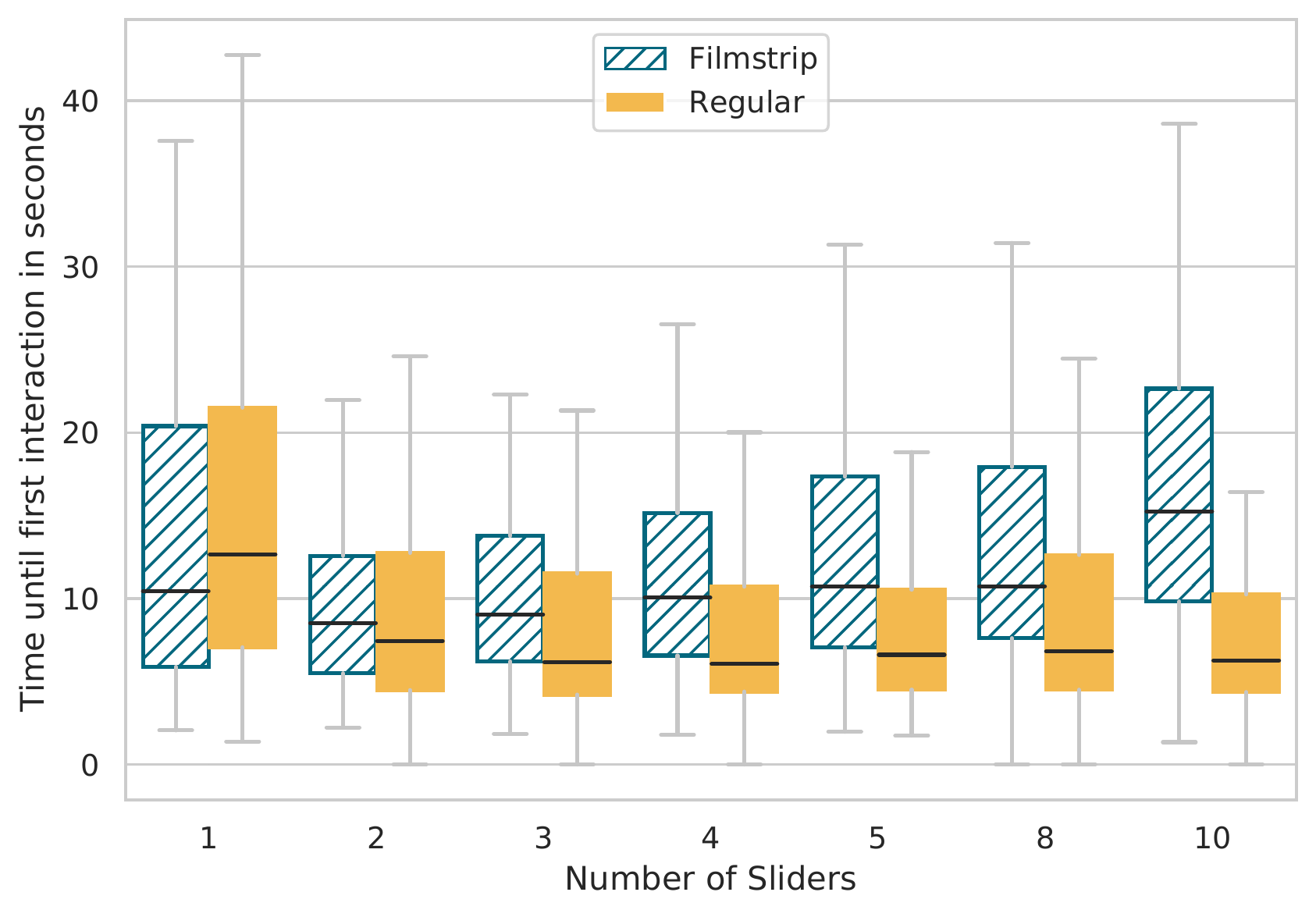}
        \subcaption{Time until the first interaction.}\label{fig:interactions_time_to_first}
        \Description{
            The median time until the first interaction is larger than 10 seconds for both variants in the setting with only one sliders. For more than one sliders, the time until the first interaction is lower than for one slider and almost constant for the regular variant. For the filmstrip variant, the time until the first interaction increases from nine seconds for two sliders to 14 seconds for ten sliders.
        }
    \end{subfigure}
    \caption{Overview of the analysed interaction metrics. All are measured per task such that each participant contributes one data point to each bar. Error bars show \pct{95} CIs.}
    \label{fig:interaction_metrics}
\end{figure*}

\subsection{Total Interactions}

To analyse the number of interactions per task (\cref{fig:interactions_number}) we counted a sequence of input events as one interaction if the user changed the same slider with at most 250\,ms between any two events (i.e. one interaction = changing one slider with a continuous drag/move or a direct click). This was informed by measuring event frequencies when dragging in various browsers as well as analysing the histogram of logged inter-event times. %

Descriptively (\cref{fig:interactions_number}), users needed more interactions to solve tasks with more control dimensions, in particular for \slidertyperegular{}. In contrast, interaction counts for \slidertypefilmstrip{} did not increase beyond 4 dimensions.
To test this effect, we fitted an LMM (Poisson family) on the interaction counts. 
The effect of \ivnumber{} was significant and positive (\glmmci{.18}{.001}{.18}{.18}{<.0001}): Each additional slider was estimated by the model to significantly increase the number of interactions by \pct{19} (computed as $\exp(\beta)=\exp(.18)=1.19$).
The effect of \ivslidertype{} was significant and negative (\glmmci{-.18}{.01}{-.21}{-.16}{<.0001}): Thus, using \slidertypefilmstrip{} instead of \slidertyperegular{} was estimated by the model to significantly reduce the number of interactions by \pct{17} ($1-\exp(\beta)$).  %
The interaction of \ivnumber{} and \ivslidertype{} was also significant and negative (\glmmci{-.11}{.002}{-.11}{-.10}{<.0001}), confirming the visible trend in \cref{fig:interactions_number} that the difference between the slider types grows with the number of control dimensions.

\subsection{Control Dimension Switches (Slider Switches)}

We counted a switch between control dimensions every time an input event modified another slider than the slider modified by the last event. 
Descriptively (\cref{fig:interactions_slider_switches}), users switched more between control dimensions if more were available, which can be combinatorially expected. \slidertypefilmstrip{} had generally fewer switches than \slidertyperegular{}.
An LMM (Poisson family) confirmed this picture:
The model had \ivnumber{} as a significant positive predictor (\glmmci{.24}{.004}{.24}{.25}{<.0001}). Thus, each additional slider was estimated by the model to significantly increase the number of switches by \pct{27}.
\ivslidertype{} was also significant (\glmmci{.08}{.04}{.00}{.16}{<.05}) and so was the interaction effect (\glmmci{-.06}{.04}{-.07}{-.05}{<.05}), which is in line with the trend visible in \cref{fig:interactions_slider_switches}: Switches increase more per control dimension for \slidertyperegular{} than for \slidertypefilmstrip{}.

\subsection{Overshooting along Control Dimensions}
We counted interactions that result in moving a slider across its target value (either in +/- direction). This measures ``overshooting'' of the target value of a control dimension in a task. %
Descriptively (\cref{fig:interactions_overshoots}), overshooting happened more often if users had more dimensions to deal with. Moreover, we observed more overshooting with \slidertyperegular{} than \slidertypefilmstrip{}. 
This is in line with the LMM (Poisson family), %
which had \ivnumber{} as a significant positive predictor (\glmmci{.17}{.003}{.16}{.18}{<.0001}): Each additional control dimension was estimated by the model to significantly increase the number of overshooting actions by \pct{23}.
\ivslidertype{} was significant and negative (\glmmci{-.33}{.04}{-.41}{-.25}{<.0001}): Using \slidertypefilmstrip{} instead of \slidertyperegular{} was estimated by the model to significantly reduce the number of overshooting actions by \pct{28}. The interaction was also significant (\glmmci{-.11}{.006}{-.13}{-.10}{<.0001}): As seen in \cref{fig:interactions_overshoots}, overshooting actions increase more per control dimension for \slidertyperegular{} than for \slidertypefilmstrip{}.

\subsection{Time to First Interaction}
We analysed the time from the start of a task to the first interaction, which might be seen as an ``orientation'' time.
Descriptively (\cref{fig:interactions_time_to_first}), this time was rather stable across the number of dimensions for \slidertyperegular{} and increasing for \slidertypefilmstrip{}. Noticeable high values appear for tasks with one slider, explained by the fact that these were the first tasks with each slider type and people thus likely looked at the task description and ``new'' UI in more detail before starting.
We fitted an LMM on this data yet found no statistically significant effects (also when excluding the first tasks).

\begin{figure*}
    \centering
    \includegraphics[width=\linewidth]{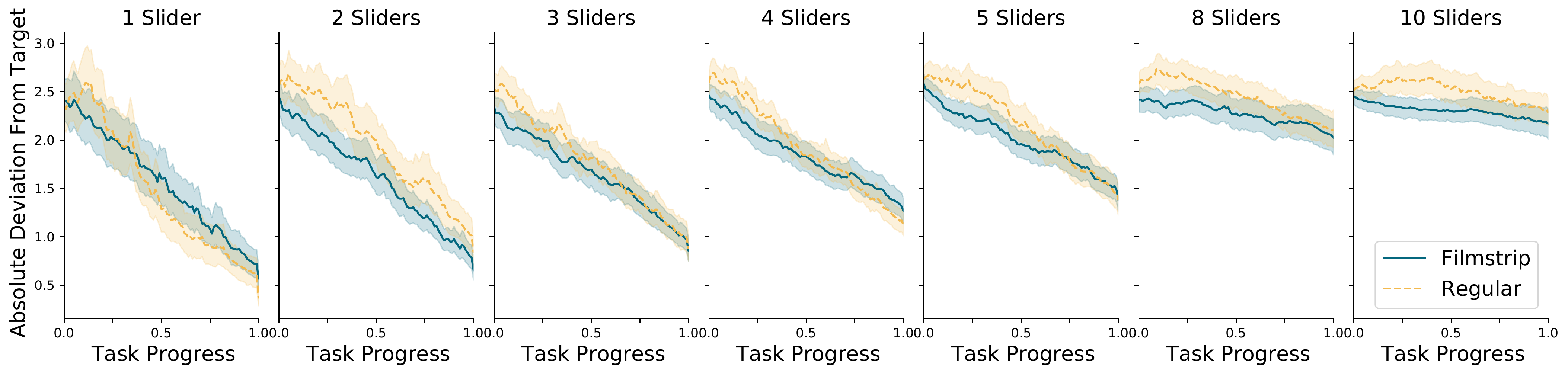}
    \caption{Absolute distance to the target over the course of interaction, averaged across participants (95\,\% CI shaded).}
    \label{fig:distance_over_time}
    \Description{
     The absolute deviation from the target decreases monotonically with increasing task progress. This decrease is less steep for increasing numbers of sliders.
    }
\end{figure*}

\subsection{Progress Patterns: Task Performance over the Course of Interaction}

Here we analyse the distance between the current slider positions and the target slider positions over the course of a task. We normalize this by the number of sliders to enable comparisons across the tasks. \cref{fig:distance_over_time} plots these distances per percent of interactions in a task (i.e. \pct{0} = first interaction; \pct{100} = last interaction), to look at progress patterns independent of absolute task time or number of interactions.
Progress is slower for tasks with more control dimensions (i.e. compare ``slopes'' across plots in \cref{fig:distance_over_time}). Moreover, \slidertypefilmstrip{} sliders tend to stay slightly closer to the targets throughout most interactions. The final distances (i.e. task accuracy) are examined in more detail in the next section. 

A subtle but noteworthy pattern here is that \slidertyperegular{} sliders move away from the target in the initial stages of a task. Closer examination of individual task logs indicates that this is because users initially move sliders seemingly randomly to grasp the effects of each control dimension. In contrast, the visual previews address this to the extent that this pattern is not visible for \slidertypefilmstrip{} sliders.

\begin{figure}
    \centering
    \begin{subfigure}{\minof{\columnwidth}{0.43\textwidth}}
        \centering
        \includegraphics[width=\linewidth]{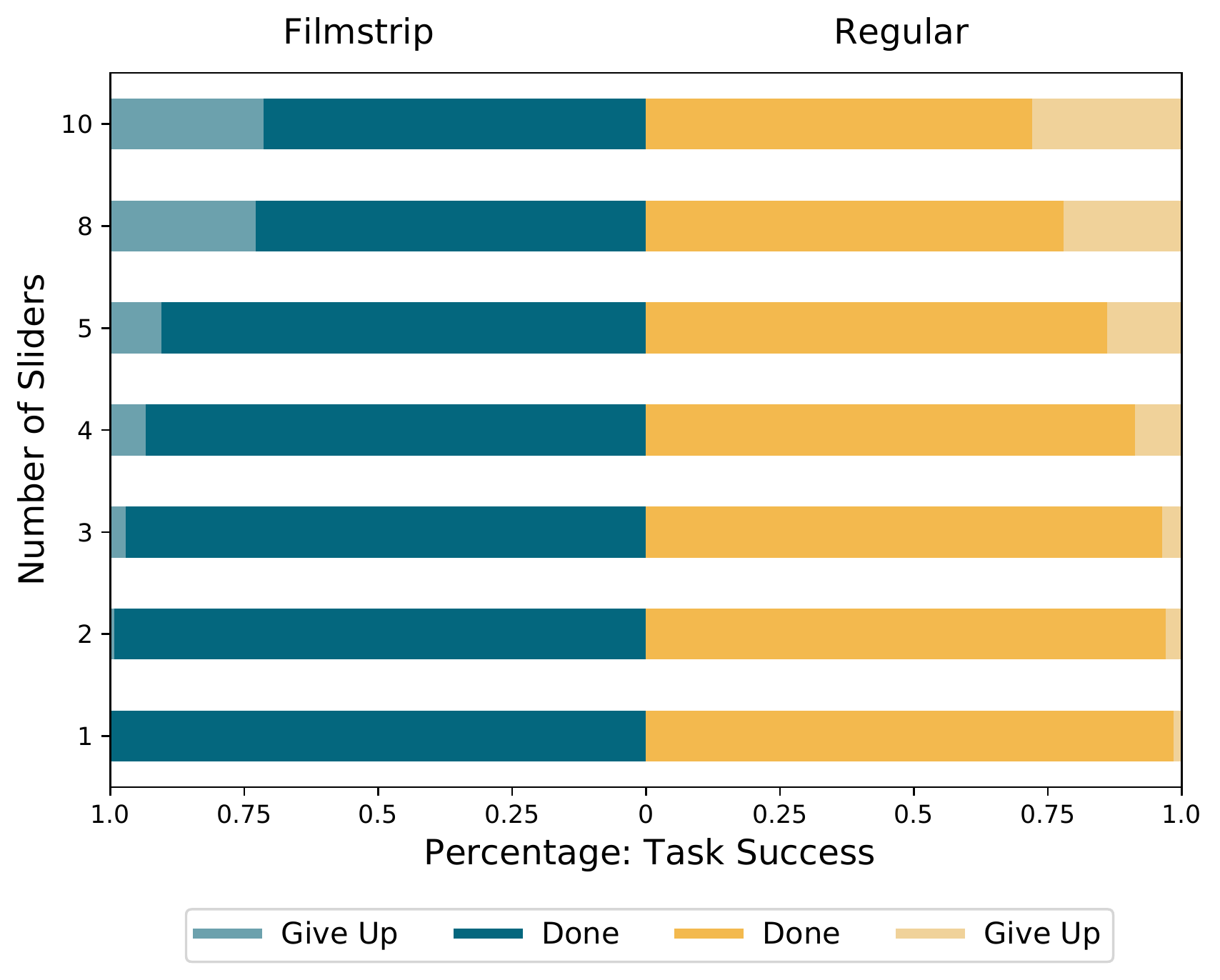}
        \subcaption{Task success}\label{fig:user_success}
        \Description{
            Overview of task success rates. The task success decreases with increasing numbers of sliders. There is no significant difference between filmstrip and regular sliders.
        }
    \end{subfigure}
    \begin{subfigure}{\minof{\columnwidth}{0.45\textwidth}}
        \centering
        \includegraphics[width=\linewidth]{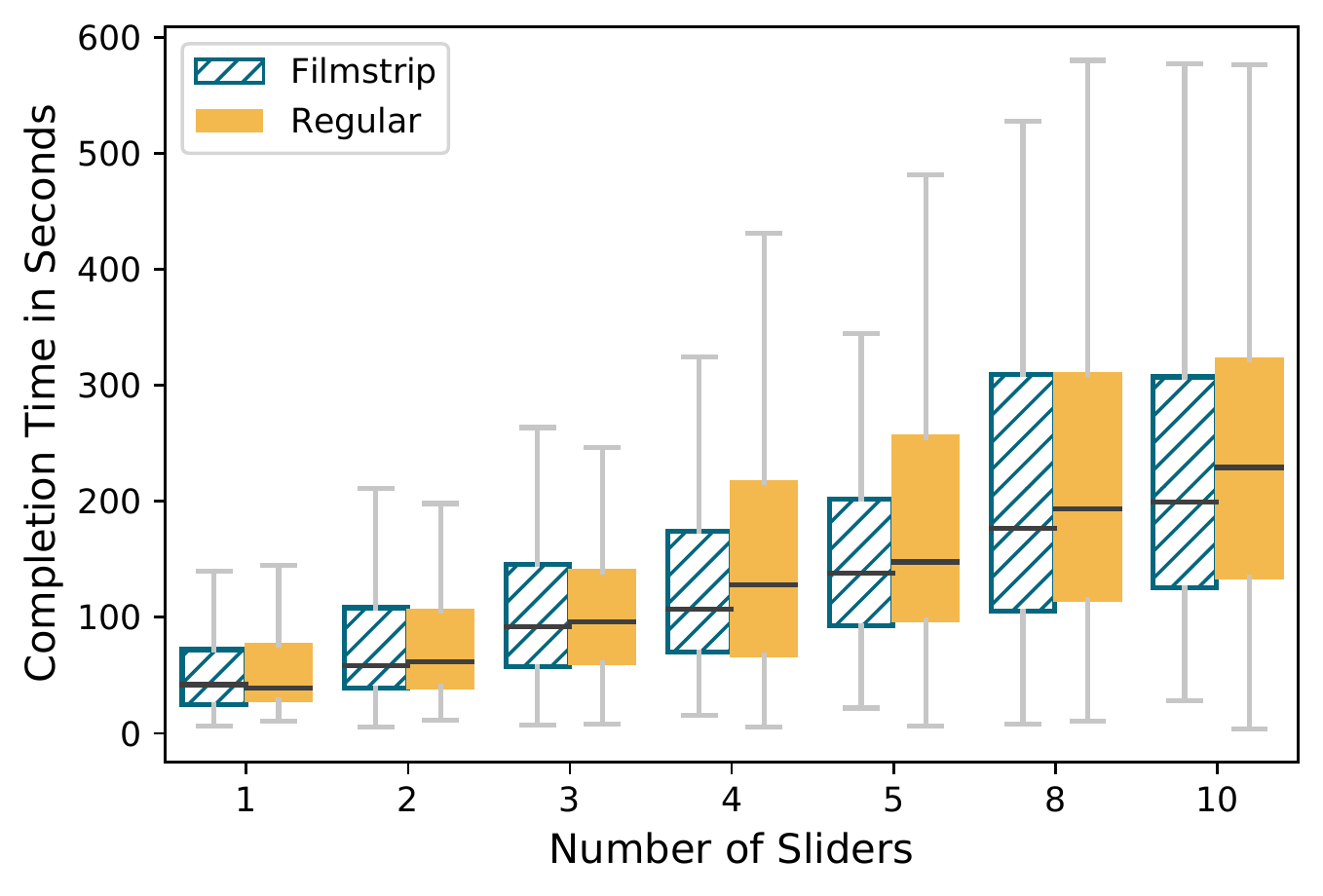}
        \subcaption{Task completion times}\label{fig:completion_times}
        \Description{
            Median task completion times increase with number of sliders. The median completion time for filmstrip is slightly lower than for regular slider. This difference is not significant.
        }
    \end{subfigure}
    \caption{Participants' task performances in terms of success and speed.}
    \label{fig:success}
\end{figure}

\subsection{Image Reconstruction Accuracy}
We measured two distances to quantify users' accuracy at the end of a task: (1) the distance between final slider positions and the task's target (i.e. ``perfect'') slider positions; %
and (2) the distance between reconstruction and target image as measured by a face recognition model~\cite{parkhi2015deep}, as a proxy for visual difference. %
We fitted LMMs for both distances: \ivnumber{} was a significant positive predictor in both (slider distance: \glmmci{2.79}{.05}{2.69}{2.90}{<.0001}; face distance: \glmmci{.04}{0.001}{.04}{.04}{<.0001}), while \ivslidertype{} was not significant for either. The interaction effect was negative for both distances yet only significant for the slider distance (\glmmci{-.20}{.08}{-.35}{-.05}{<.01}).
In summary, in line with the trends visible in \cref{fig:distance_over_time}, people solved the tasks significantly less accurately for more control dimensions, and there is a trend that the filmstrip design overall slightly mitigated this decrease in accuracy.

\subsection{Task Completion Success}

People could end a task either by marking it as ``done'' or ``skip'' (\cref{sec:ending_a_task}). %
\cref{fig:user_success} shows this data. An LMM (binomial family) had \ivnumber{} as a significant negative predictor (\glmmci{-.36}{.04}{-.44}{-.28}{<.0001}): Tasks with more control dimensions had a significantly lower chance of success. %
However, absolute rates of people's indicated success were rather high. We did not find a significant effect of \ivslidertype{} or interaction effect. Moreover, face recognition distances were significantly higher when giving up (\ttest{-.27}{-.30}{-.25}{266.77}{-23.34}{<.0001}{-1.70}), underlining that people used these choices meaningfully.%

\subsection{Task Completion Time}
We measured task completion time (\cref{fig:completion_times}) from the start of a task to submitting it as ``done''. %
An LMM on this data (in ms) had \ivnumber{} as a significant positive predictor (\glmmci{22535}{1497}{19601}{25468}{<.0001}): Tasks with more control dimensions took significantly longer. \ivslidertype{} was a negative predictor (i.e. \slidertypefilmstrip{} reducing times) but not significant. We also did not find a significant interaction effect.

\begin{figure*}
    \centering
    \includegraphics[width=\linewidth]{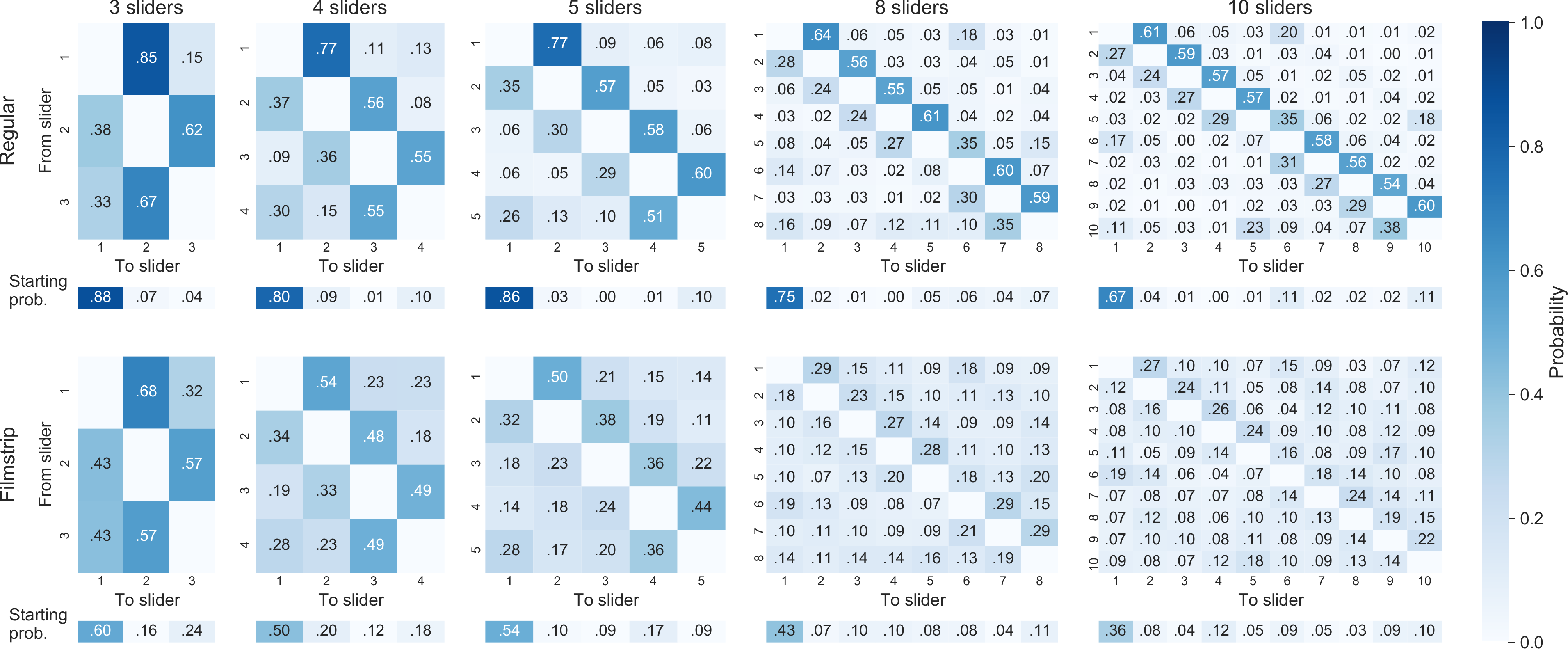}
    \caption{Overview of users' transitions between different sliders during interaction. For each slider number and design,  %
    the matrix plot shows the transition probabilities of switching from one slider (row) to another (column) during interaction (i.e. rows sum to 1). %
    This is non-trivial only for >2 sliders. %
    The single row plots show the starting probabilities per slider in each task%
    . See text for details.}
    \label{fig:sequences}
    \Description{
        The overview shows the users' transitions between different sliders during interaction. For the regular slider variant, we observe a sequential interaction pattern (i.e. users transition from slider 1 to slider 2, then from slider 2 to slider 3, and so on. This pattern is not observed for the filmstrip slider.
    }
\end{figure*}

\subsection{Exploration Behaviour}\label{sec:exploration_behaviour}

We analyzed the interaction logs as sequences: Concretely, \cref{fig:sequences} shows how often users transitioned between specific dimensions (e.g. row 1, column 3 = changing the 1st slider, then the 3rd). These plots reveal differing strategies per slider design: For \slidertyperegular{}, people started at the top of the UI (i.e. slider 1, see starting probabilities in \cref{fig:sequences}) and tried out sliders one after the other (i.e. highest values just above the diagonals in \cref{fig:sequences}). They sometimes also went back up (i.e. rather high values just below the diagonals in \cref{fig:sequences}). This sequential pattern is far less pronounced for \slidertypefilmstrip{}, in particular with increasing control dimensions. While people here also tend to start at the top of the UI, transitions are much more varied, indicating that they used the visual information to decide which slider to change next.

\subsection{Task Perception and Subjective Feedback}

Here we report on the results from the questionnaires and open feedback (see \cref{sec:study_questionnaires}).

\begin{figure*}
    \centering
    \includegraphics[width=\textwidth]{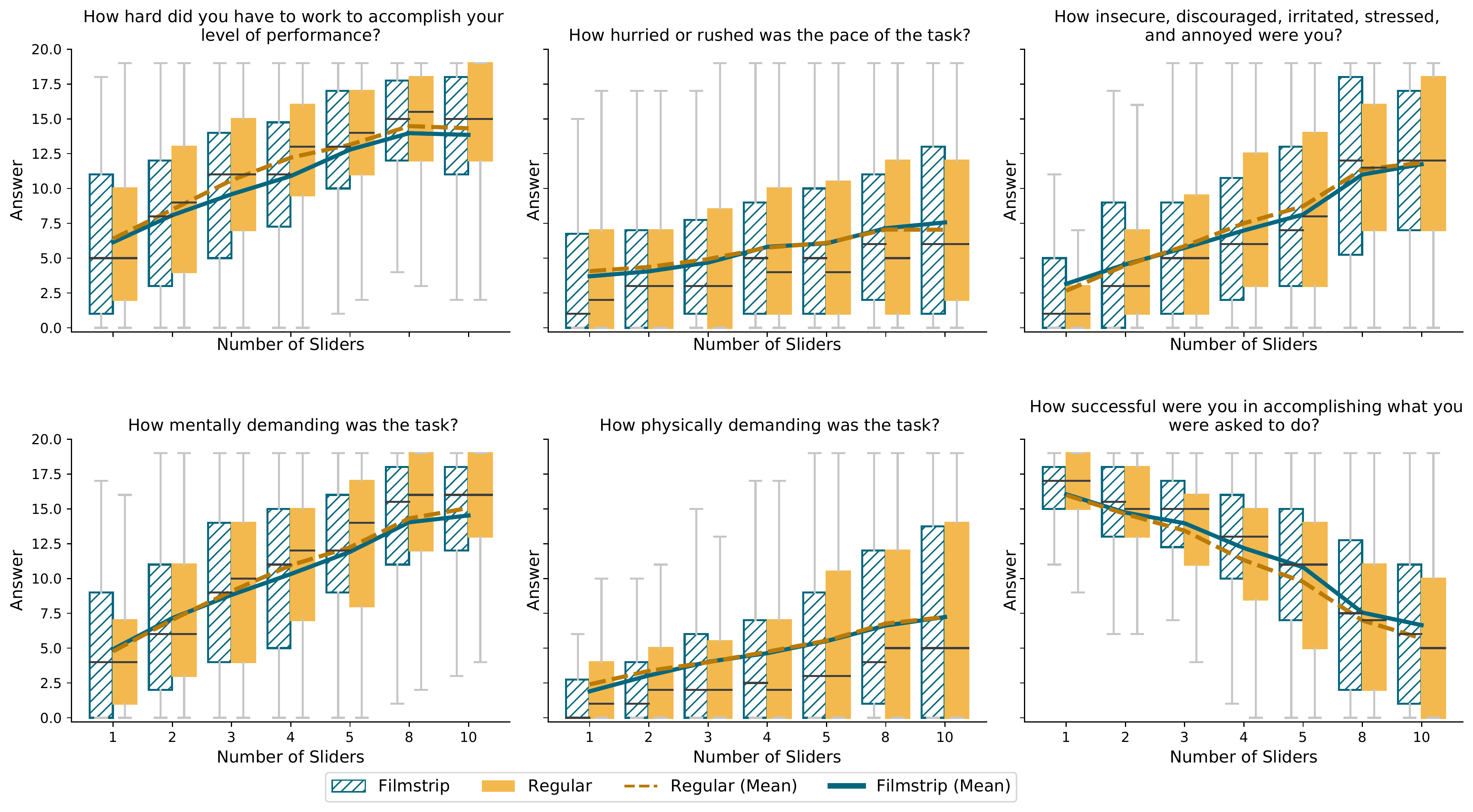}
    \caption{Overview of user responses to NASA-TLX questionnaire. There are six dimensions. Overall, the perceived workload increases with increasing numbers of sliders. The perceived success decreases with increasing numbers of sliders.}
    \Description{Figure shows six plots, each corresponding to one dimension defined in NASA-TLX. Five out of six plots indicate a increasing trends towards more perceived workload with increasing numbers of sliders. One plot indicates decreasing preceived success with increasing numbers of sliders.}
    \label{fig:TLX}

\end{figure*}

\subsubsection{Task Load Index}
Figure \ref{fig:TLX} shows participants' NASA-TLX ratings, which we analyse on subscale level~\cite{Hart2006}: 
Concretely, the LMM analysis showed that \ivnumber{} was a significant predictor for all NASA-TLX facets (mental demand: \glmmci{1.08}{.05}{.99}{1.17}{<.0001}; physical demand: \glmmci{.47}{.04}{.40}{.53}{<.0001}; effort \glmmci{.88}{.05}{.79}{.97}{<.0001}; temporal demand: \glmmci{.36}{.04}{.28}{.43}{<.0001}; performance: \glmmci{-1.00}{.06}{-1.09}{-.91}{<.0001}; frustration: \glmmci{.95}{.05}{.87}{1.04}{<.0001}). Thus, people perceived tasks with more sliders as significantly more demanding and frustrating and as requiring significantly more effort. They also perceived their own task performance as significantly lower.
Neither \ivslidertype{} nor the interaction of \ivnumber{} and \ivslidertype{} was found as significant for any of the questions.

\subsubsection{Likert Items per Task}

\begin{figure*}
    \centering
    \includegraphics[width=\linewidth]{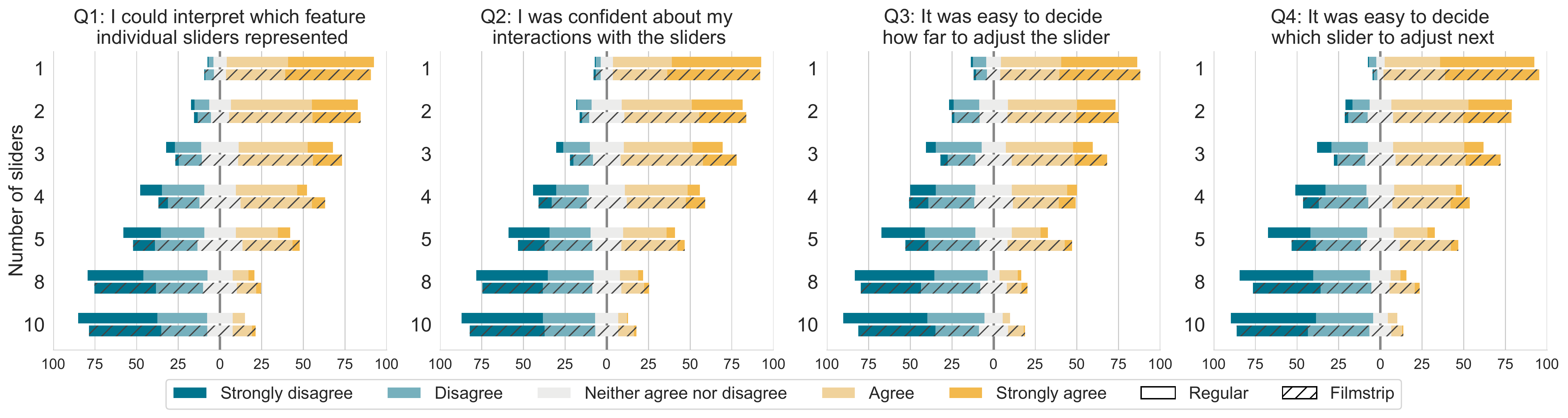}
    \caption{Overview of Likert results from the questionnaires after each task (x-axes: percent of respondents). Overall, tasks with more sliders received worse ratings and tasks using the Filmstrip visualizations were rated slightly better. }
    \label{fig:likert}
    \Description{
     Plot shows an overview of the Likert results from the quesstionnaires after each task. The titles of the four plots read: (Q1) I could interpret which feature individual sliders represented, (Q2) I was confident about my interactions with the sliders, (Q3) it was easy to decide how far to adjust the slider, (Q4) it was easy to decide which slider to ajdust next. The filmstrip sliders received slightly better user responses. Tasks with more sliders received worse ratings.
    }
\end{figure*}

\cref{fig:likert} shows the results of the Likert items after each task assessing the sliders' interpretability, user confidence, and ease of interaction. Descriptively, tasks with more sliders received worse ratings and tasks with \slidertypefilmstrip{} sliders received marginally better ones. The results from the GEE analysis match this picture: \ivslidertype{} was a positive predictor, yet not significant, while \ivnumber{} was a significant negative predictor for all questions.
Concretely, the odds of giving a higher Likert rating decrease with an additional slider: They are estimated by the model as X times the odds without that slider (with X=.66 for $Q_1$, .64 for $Q_2$, .66 for $Q_3$, and .65 for $Q_4$; all p<.0001). This means that having more sliders results in a perception of: (1) significantly worse interpretability of individual sliders; (2) significantly lower confidence in one's interactions; (3) significantly harder decisions on how far to adjust the sliders; and (4) significantly harder decisions on which slider to adjust next.

\subsubsection{Open Feedback}\label{sec:open_feedback}

The study concluded with three open questions (see \cref{sec:study_final_questionnaire}).
Two authors developed a codebook inductively and then coded all responses independently. All differences were resolved through discussion. 

\textit{Interpretability of the slider types:} 51 people provided comments on this question. Many compared the sliders. For example, one person in favour of \slidertypefilmstrip{} explained: \textit{``Sliders with photos stitched together were definitely better than those without. I enjoyed working with them and I had fun. Working with sliders without photos was very annoying and stressful.''}
In contrast, another person said: \textit{``I believe the plain slider as opposed to the filmstrip slider was much easier to interpret as the user is able to focus on the main two images rather than getting distracted by the images within the filmstrip slider, this allows them I think to be more accurate and focused.''}
Overall, \slidertypefilmstrip{} was preferred here: \pct{31} explicitly mentioned something positive about \slidertypefilmstrip{} vs \pct{6} about \slidertyperegular{}; and \pct{37} explicitly expressed a preference for \slidertypefilmstrip{} vs \pct{8} for \slidertyperegular{}. 
Beyond such comparisons, further mentioned aspects included comments on: slider numbers (\pct{14}), mostly stating that more sliders make it more difficult; hard to interpret dimensions (\pct{12}); and performance, such as lags (\pct{8}).

\textit{Ideas for UI improvements:} 78 participants responded to this question. Here, the top aspects were: adding slider labels (\pct{35}); improving performance (\pct{18}); and adding UI elements (\pct{17}), such as a reset or undo button or an option to save the current state. Further suggestions included enabling more fine-grained control (\pct{8}) and having more meaningful dimensions (\pct{6}). We further analyzed the aspects mentioned as potential labels for such systems: \textit{demographics} (e.g. age, perceived gender, ethnicity), \textit{appearance} (e.g. facial features like eyes, nose, mouth and hair, or accessories such as jewelry or glasses) and \textit{image properties} (e.g. hue, saturation, contrast, brightness).

\textit{Potential other use cases:} 57 people answered this question. The top aspects were: image editing (\pct{25}); criminology (\pct{26}); prediction (\pct{21}) such as on ageing, hairstyles, glasses or jewelry; and gaming (\pct{7}). %

%% file: sections/discussion.tex
\section{Discussion}\label{sec:discussion}

\subsection{Measuring and Addressing the Costs of Dimensionality in Controlling Generative Models}

As a key result, we provide the first quantification of interaction costs per control dimension for a generative model for images: Each control dimension, operationalised in the UI as a slider, is estimated to increase number of user actions towards a desired outcome by +\pct{19}. It also increases workload.
More dimensions also make interaction paths towards the target image more complex, with an estimated +\pct{27} increase of switches between dimensions per additional slider. 
\citet{Ross2021} recently shared the expectation that ``interacting with all dimensions simultaneously may be overwhelming for models with many tens or hundreds of representation dimensions''. Our analyses and people's feedback here indicate that this can become challenging already with about five to ten dimensions.

Based on these results, end user applications should carefully consider how many control dimensions to simultaneously show as sliders. Combined with people's comments on interpretability and the Likert ratings, we recommend at most 3-5. This is likely a conservative estimate, in the sense that dimensions with a higher degree of interpretability~\cite{Ross2021} can be expected to impose lower interaction costs. In this light, our results provide an additional, interaction-centric motivation for work on interpretability: Reducing interaction costs for UIs with multiple control dimensions.

In contrast, research applications might need UI controls for many dimensions, such as for exploring a generative model in development. Here, our results motivate %
UI concepts that let people externalise their exploration of said (potentially many) dimensions. %
Beyond allowing for entering text labels~\cite{ganspace, Ross2021}, our insights motivate exploring concepts from infovis and information retrieval (e.g. searching, reordering, ranking, filtering or selecting control dimensions). For example: (1) users could hide sliders in the UI if they deem them to be currently uninteresting; (2) users could group or merge sliders if they identify them as similar or more meaningful as a single interactive control; (3) or sliders could be reordered in the UI, either manually or automatically according to various (selectable) metrics. %

\subsection{Feedforward Information for Interactive Generative Models for Images}
Here we discuss the observed benefits and challenges around feedforward information in this context.

\subsubsection{Feedforward Information Facilitates Selective and Strategic Control of Generative Models}
A clear benefit of the examined slider design with visual previews is that it mitigates the costs of dimensionality described in the previous section: That is, with preview images on the sliders, the required number of interactions tapers off at about three control dimensions, instead of increasing further (\cref{fig:interactions_number}). Similarly, the previews also reduce the number of slider switches (\cref{fig:interactions_slider_switches}) and overshoots (\cref{fig:interactions_overshoots}). That is, they reduce patterns indicating uncertainty such as ``jumping'' back and forth between dimensions, and along dimensions, instead of setting them correctly once. Moreover, the previews facilitate strategies beyond simply going by the order of the sliders in the UI layout (\cref{fig:sequences}). Related, some people commented that the previews helped with interpretation, adjustments and avoiding ``trial and error''.

To interpret this further, we can view image reconstruction as a search problem in which users have to find the target image within the part of the latent space that is interactively accessible through the sliders. In this view, adding a dimension expands the space exponentially (e.g. $n^d$ images for $d$ sliders with $n$ discrete values each). As our results show, users' strategies allow them to avoid an exponential growth in their required actions: With regular sliders, users achieve linear growth, that is, adding a slider adds a roughly constant number of required interactions (\cref{fig:interactions_number}). Visual feedforwad allows them to achieve a logarithmic pattern instead (i.e. diminishing costs per slider).

Overall, these findings support the conclusion that visual feedforward information in the goal-directed control of a generative model for images leads to more selective and strategic decision-making throughout the interaction. %

\subsubsection{No Evidence for Improved Image Reconstruction Performance with Feedforward}
Our findings do not support the conclusion that the strategic benefits of the filmstrips described above result in faster or more accurate reconstruction: First, we did not find a significant effect on task completion times. A likely explanation is that users needed extra time to attend to the previews. Preview updates also had a short delay. Moreover, a few people commented that they felt distracted by the previews. Related, previews seemed to invite spending more time on orientation before the first interaction (\cref{fig:interactions_time_to_first}) although this was not statistically significant.

Second, we could not find a significant effect on the final reconstruction accuracy: %
An explanation is that, for both slider designs, people found strategies to solve the tasks, and difficulty was rather mostly dependant on the number of control dimensions. This is supported by the interaction measures and patterns, revealing different strategies for the two slider types. While these strategies did not lead to significantly different perceived workload and performance (\cref{fig:TLX} and \cref{fig:likert}), the open feedback was clearly in favour of the filmstrip design (\cref{sec:open_feedback}). %

\subsubsection{Summary and Ideas for Future Work}
Overall, our takeaway is to use and further explore visual feedforward for interactive control of generative models: While people can indeed solve image reconstruction tasks with blank sliders, the visual feedforward information facilitates more goal-directed interactions. This is also subjectively perceived as such by many users, resulting in a positive and preferable user experience overall. 
Finally, as methodological guidance, researchers studying generative models via interaction need to be aware that the regular slider design mainly elicits a simple sequential user strategy (i.e. going through the dimensions as presented in the UI layout).

Future studies could explore further designs in comparison to our fundamental filmstrip design here. For example, further ideas motivated by our results and participants' suggestions include showing only two images per slider (e.g. at min and max state) or ``derivative'' information (e.g. image showing difference between max and min state). Such designs could also be compared for models of varying disentanglement, effectively motivating a study that combines the key independent variables of this paper and those in the work by \citet{Ross2021}. Moreover, sliders could be compared to different UI concepts in this context (e.g. image grids~\cite{Zhang2021}).

\subsection{Quantifying Interactions with Generative Models beyond Time and Accuracy}
We found that it is crucial to study interaction with generative models beyond the typical HCI performance metrics of task time and accuracy: The two slider designs were similar in these measures despite drastic differences in users' behaviour and strategies. These were only revealed by detailed interaction metrics (\cref{fig:interaction_metrics}, \cref{fig:sequences}). Recent related work has proposed performance measures for interaction with generative models~\cite{Rafner2020chiplay, Ross2021}, namely completion rate and time, error AUC (a summary of what we plot over time in \cref{fig:distance_over_time}), and self-reported difficulty. We contribute further metrics, including: number of interactions, number of switches between control dimensions and overshooting within dimensions, and starting/transition probabilities between control elements in the UI.
Studies of interactive control of generative models have only recently appeared in the community and called for further explorations~\cite{Rafner2020chiplay, Ross2021, Zhang2021}. In this light, we hope these metrics provide timely methodological inspiration. To further facilitate reuse, we release the code to compute these metrics as well as our dataset as a point of comparison for future measurements. %

\subsection{Discovering Interpretations of the Control Dimensions}

People were overall successful in completing the tasks but in the open feedback wished for explicit labeling of sliders:
Here, we found demographics, appearance and image properties as the three main label categories (\cref{sec:open_feedback}). This shows an interest in fine-grained control over several aspects of the image. %

We conclude that visual feedforward helps users to better predict which slider to adjust and how (\cref{sec:exploration_behaviour}). However, it is insufficient for in-depth understanding of the underlying dimensions. Using the terms of recent work on understanding of intelligent systems~\cite{Eiband2021}, previews mainly facilitated \textit{interaction knowledge}: Users learned how to use the system effectively -- without necessarily being able to explicitly explain it. Here, text labels may help yet require an interpretable model. Note how in the presence of labels previews could fill a new role: While labels would introduce the broad concept of a slider the previews would give an impression of the concrete impact. In other cases, or if labels are not available upfront (e.g. in research~\cite{ganspace}), our results motivate a combination: Previews could help users to identify and label relevant dimensions for further use. %

\subsection{Limitations}\label{sec:limitations}

We studied a general user population here. Future studies could investigate specific groups (e.g. AI researchers).

Preview images are computationally costly. We switched to a multi-GPU server after a pretest to handle this, and observed times for updating all previews of about 700\,ms to 2500\,ms, depending on the number of sliders. About \pct{10} of all users commented on performance. We deem this acceptable for an online study in which we cannot control people's internet connection and computing environment. When interpreting the results it should be taken into account that this might have influenced the comparisons of task times, perception and success. However, task times were still clearly dominated by the difficulty of tasks with more sliders, as evident from the increased number of actions (\cref{fig:interaction_metrics}), a trend towards slightly faster times with filmstrips (\cref{fig:completion_times}) and people's feedback. Moreover, we found no significant differences between the designs in the NASA-TLX and success rates. Finally, our UI updating strategy (\cref{sec:technical_implementation}) ensured that the sliders were still usable during updates. %
Speed-ups might be achieved with decreasing model sizes or other optimisations. However, changing a studied model needs to be carefully considered in light of a study's goals. The chosen dataset may also influence the image resolution required to visually discern thumbnails and thus indirectly influence the render speed of the system. %

Generalisation of our results may be limited due to the dataset (faces), model (\textit{StyleGAN2}), automatic extraction of dimensions (\textit{GANSpace}) and UI concept (sliders). For example, distinguishing subtle changes may be easier for faces than cars, and handcrafted dimensions may be easier to interpret and control.
However, we expect similar effects of the number of individually controlled dimensions and visual feedforward for other generative models with visual output.
Future studies could examine such effects for other datasets (i.e. beyond faces), other media types and feedforward modalities (e.g. text, audio), and other models.

\subsection{Broader Reflections}

Generative models and their applications have been criticised for biases~\cite{vincent_what_2020}. Our study adds the following insights:
\citet{Matejka2016} found that visual markers on sliders influence the result of simple value choices. We revealed a related effect for the control of generative models: Preview images change users' strategies and therefore the outputs seen during interaction. %
In principle, it might thus be possible to design UI controls in a way that favours users' exposition to certain outputs when interacting with the model. %
Moreover, related work shows evidence that visual slider midpoints are perceived as a conceptual center or most typical value~\cite{tourangeau_spacing_2004}. Therefore, model output shown as previews on a slider might be implicitly presented as ``normal'' (center) or not (ends). 
Overall, our work highlights that applications of generative models need to look also beyond models to address bias: Careful design of their UI controls is also important. 

For use cases with images in particular, our study also newly emphasises the influence of the \textit{input} UI elements, complementing HCI literature that discusses bias in image \textit{outputs} (e.g. in image search results~\cite{Kay2015}).

Finally, generative models can also be used to reveal biases in other AI systems, by generating counterfactual inputs~\cite{Denton2019}: The related work here called for further human assessment of ``[...] images that result from manipulating the latent space''. This in turn requires UIs such as the ones that we studied here. 
Together, these aspects motivate further studies at the intersection of HCI and AI on interaction with generative models.

%% file: sections/conclusion.tex
\section{Conclusion}

We have investigated the impact of different numbers and types of sliders on users' interaction behaviour and performance in image reconstruction tasks with a generative model. Overall, we found that exposing users to more control dimensions steeply increases the number and complexity of interactions. However, using the filmstrip slider variant mitigates the extent of this effect. These results also refine prior expectations in the literature~\cite{Ross2021} by showing that control of generative models can get very challenging even for up to ten sliders. Adding to the methodology in this context, we found that measures beyond task completion times and overall success rate are crucial to characterize the impact of different UI designs.

Our results challenge the HCI and AI research community to further investigate the UI design space for interactive generative models: While regular sliders are frequently used, there is still room for improvement. The filmstrip slider is one step in that direction. %
We make the collected interaction dataset and prototype publicly available to facilitate research towards the vision of empowering experts and non-experts alike to work with generative models:

\url{https://osf.io/tze2x/}%

%% file: sections/appendix.tex
\onecolumn
\appendix
\section{Appendix}\label{sec:appendix}

\begin{figure*}[b!]
    \centering
    \includegraphics[height=17cm]{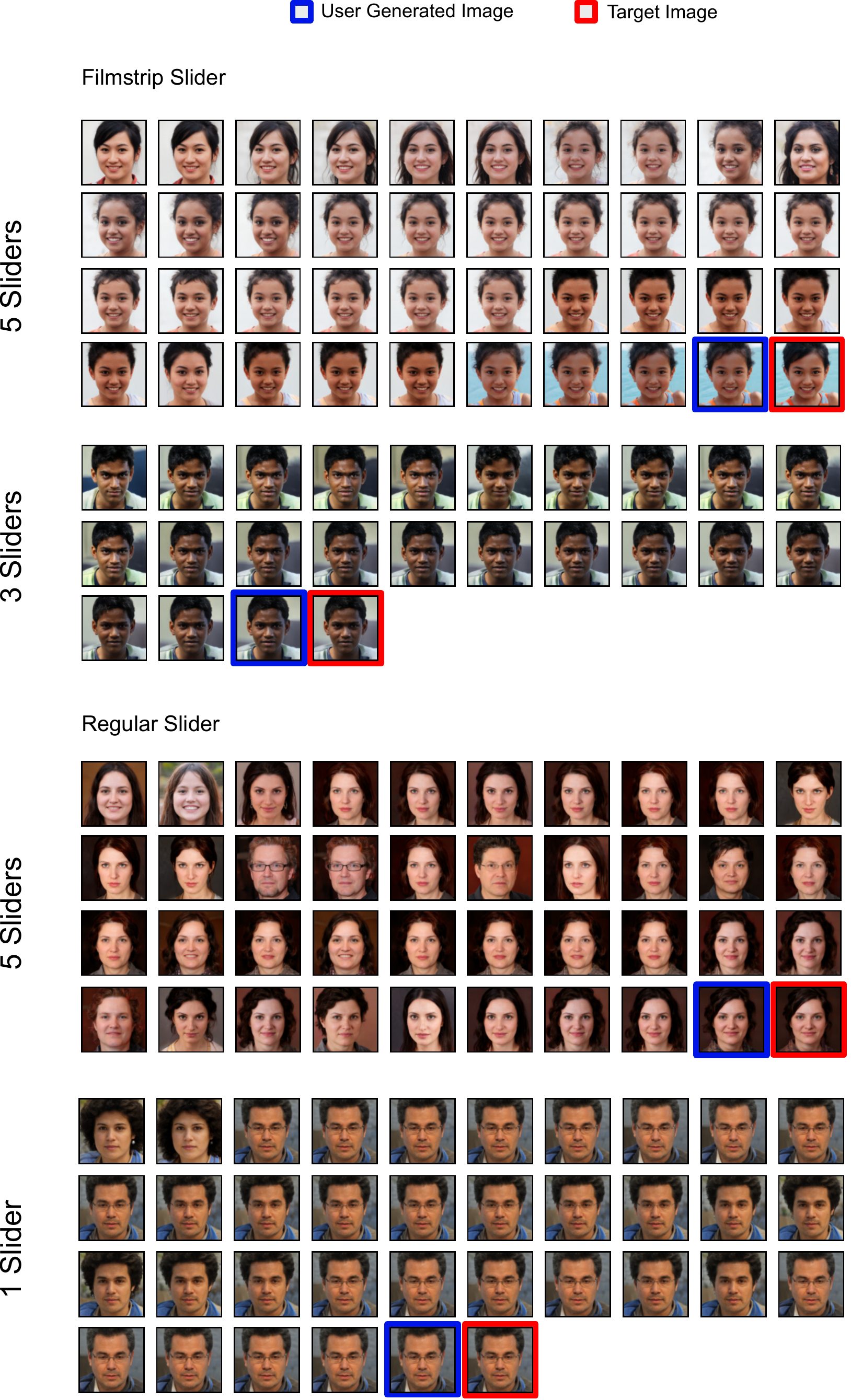}
    \caption{Example of four different user exploration paths where users marked the task as \textit{done}. The last image (red) marks the target image. The penultimate images show the final user edited images (blue). Each image represents the state after an edit. For more than 40 interactions we downsampled to 40 images for the sake of brevity of the visualization.}
    \label{fig:path_done}
 \end{figure*}

\begin{figure*}[!b]
    \centering
    \includegraphics[height=9cm]{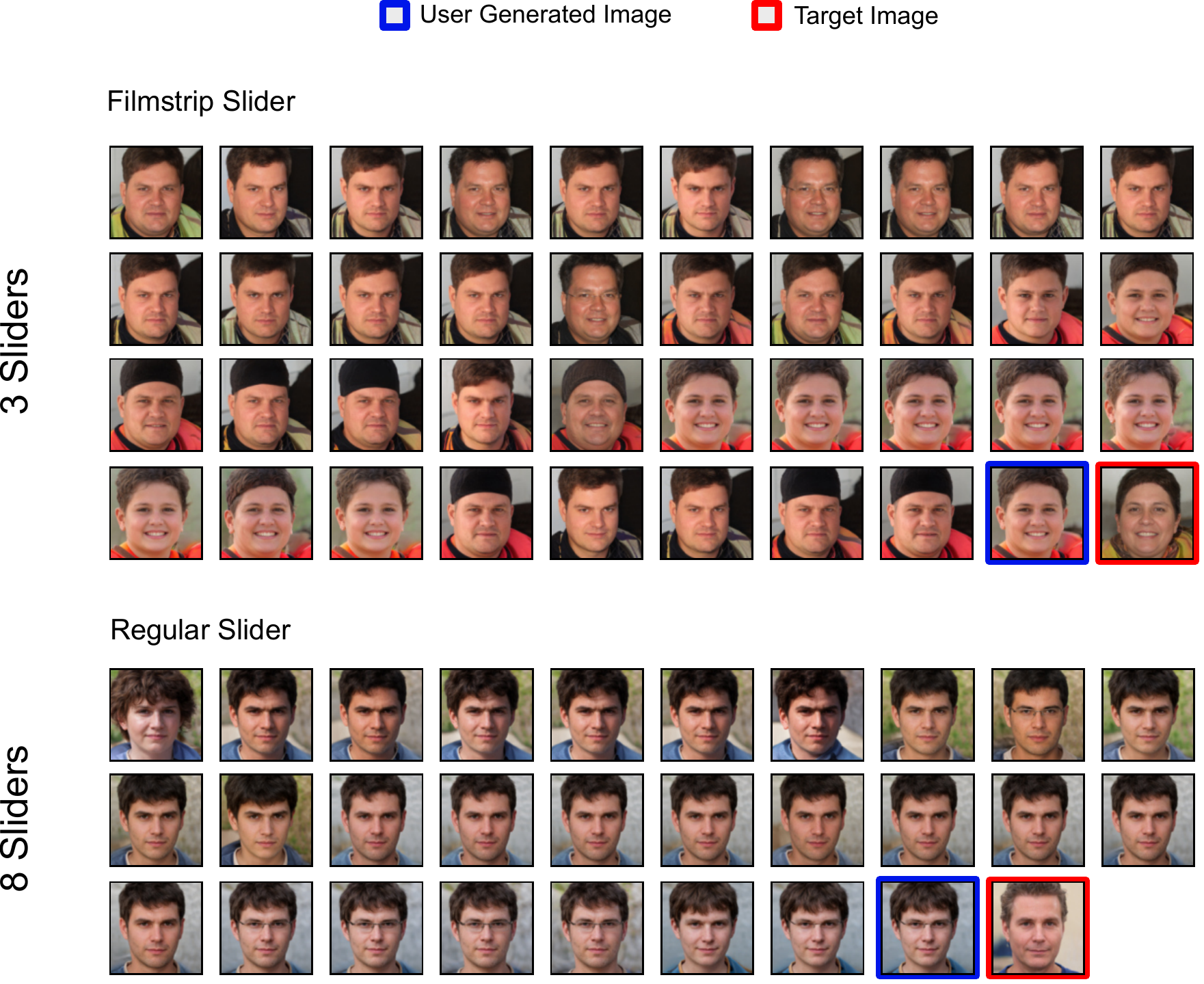}
    \caption{Example of two user exploration paths where users eventuelly indicated that they want to \textit{skip} the task. The last image (red) marks the target image. The penultimate images show the final user edited images (blue). Each image represents the state after an edit. For more than 40 interactions we downsampled to 40 images for the sake of brevity of the visualization.}
    \label{fig:paths_skip}
\end{figure*}